# A comparative study between seasonal wind speed by Fourier and Wavelet analysis

Sabyasachi Mukhopadhyay[a], Debadatta Dash[b], Asish Mitra[c], Paritosh Bhattacharya[d]

[a]Indian Instiute of Science Education Research Kolkata, Mohanpur Campus, Nadia-741252, India
[b]Veer Surendra Sai University of Technology Burla, Odisha- 768018, India
[c]College of Engineering & Management, Kolaghat, Purba Medinipur-721171, India
[d]National Institute of Technology Agartala, Agartala-799055, India

**Abstract:** Wind Energy is a useful resource for Renewable energy purpose. Wind speed plays a vital role for wind energy calculation of certain location. So, it is very much necessary to know the wind speed data characteristics. In this paper fourier and wavelet transform are applied to study the wind speed data. We have compared wind speed of winter with summer by taking their speed into account using various discrete wavelets namely Haar and Daubechies-4 (Db-4). Also the periodicity of wind speed is checked using Continuous Wavelet Transform (MCWT) like Morlet. Thereafter a comparative study is done for detecting the periodicity of both summer and winter. Then wavelet coherence is checked between these two data for extracting the phase coherency information.

**Keywords:** Fourier Transform, Wavelet Transforms, Wavelet Coherence, Scalogram, Periodogram, Wind speed.

## 1. Introduction

Wind is a principal component of our nature. Currently it is being used as a renewable source of energy. So, it is very essential to know the characteristics of wind data. For the characterization of the Wind speed and Wind energy, the role of mathematical tools is inevitable. The utility of the Hilbert Transform is shown for the low wind speed [1]. Mukhopadhyay et al., introduced the optimized DHT-RBF based analysis for Wind power forecasting purpose [2]. The discrete Hilbert Transform (DHT) has been used as a minimum phase type filter for characterizing the wind speed data by Mukhopadhyay et al.,[3]. Panigrahi et al., introduced wavelet in weather related application [5]. Mukhopadhyay et al., analysed the wind speed data of summer of Eastern region of India by the continuous wavelet transform and multifractality [6]. Liu et al., used wavelet transform based analysis in the field of Ocean technology field [8].He et al., used wavelet transform for analyzing the wind data in Dongting lake cable stayed bridge Region [8]. Wind data simulation of Saudi Arabia region was done by wavelet transform [9]. Mukhopadhyay et al., introduced s transform for the wind speed data analysis purpose [11].

## 2. Details of Wind Speed data pattern

We have taken the experimental wind speed data of East Midnapore district of west Bengal. We observed the wind speed very carefully and cautiously using anemometer. First we took the daily wind speed of December-January (2010-2011) (fig 1.(a)), in the winter season and then we observed the summer wind speed during April-may (2011)(fig 1.(b)).

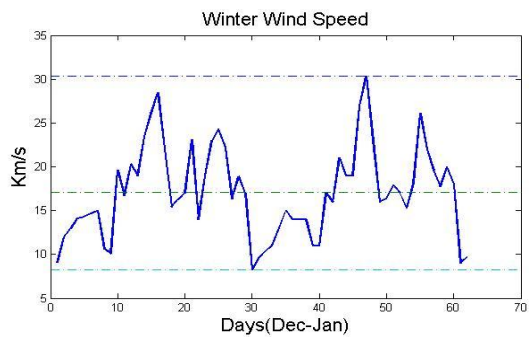  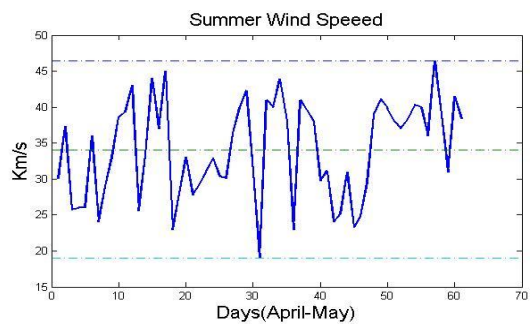

**Fig 1(a):** Winter wind speed       **Fig 1(b):** Summer Wind Speed

From the data statistics of both the seasons we see that comparatively in the winter season the wind speed is higher. We also observed during our experiment that the difference between the wind speed of this two season is much more higher than the other seasonal speed differences, which in a way explains the opposite seasonal weather conditions.

## 3. Results and Discussions

At first Fast Fourier Transform (FFT) has been applied for summer and winter wind data to observe the Fourier coefficients and their distribution in Fourier plane (Fig 2(a, b)).

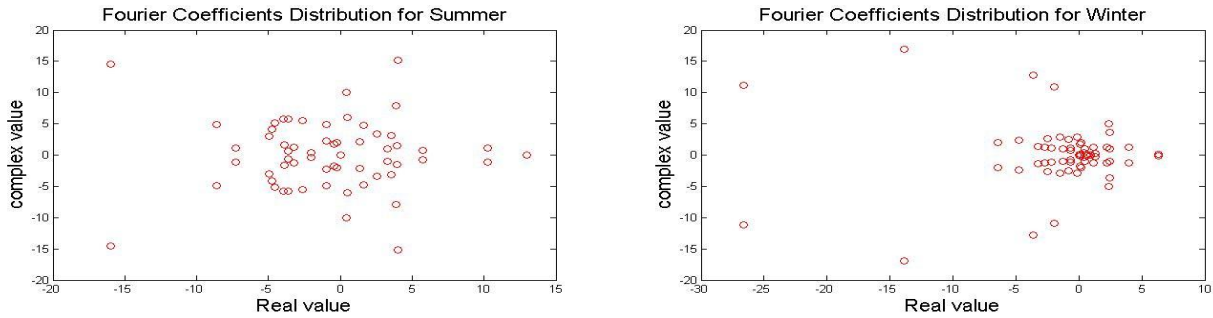

**Fig 2:** Fourier coefficients distribution plot for summer (left) and winter (right) wind speed data

From the comparison of above two figures it is clear that for winter case the Fourier coefficients are more concentrated than Fourier coefficients of summer wind data. But in case of summer, the fourier coefficients are more scattered. It actually predicts the agility of the wind particles in summer due to rapid convection process. In case of winter these convection processes are very less due to low convection process. To know the periodicity of wind speed data, Periodogram plots are done, i.e., the frequency plot of Fourier power which is the sum of square of absolute values of Fourier coefficients as shown below.

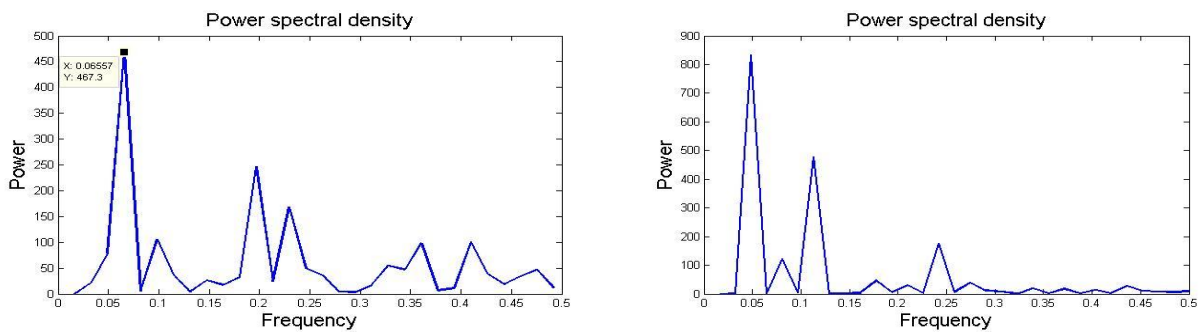

**Fig 3(a):** Periodogram of Summer wind speed Data     **Fig3(b):** Periodogram plot of winter wind speed data

From the above plots Fourier power peaks are observed with their corresponding frequencies. By inverting this frequency corresponding to maximum power, the periodicity can be approximated as 20 days for summer and 15 days for winter.
Now, to have a close look at the periodicity peaks, wavelet transform is introduced for the analysis purpose. Due to the presence of the scaling function, zooming operation can easily be done here which provide us some distinct features about summer and winter wind data.

The first and simplest wavelet to be introduced is HAAR which is nothing but a step function satisfying the wavelet conditions. This is a discontinuous wavelet and same as Daubechies-1, i.e., db1. At first the simple HAAR wavelet is chosen to study and differentiate the summer wind speed from winter wind speed. So both of the data are decomposed by HAAR wavelet up to $5^{th}$ level using 'WAVEMENU' interface in MATLAB command prompt and plotted them.

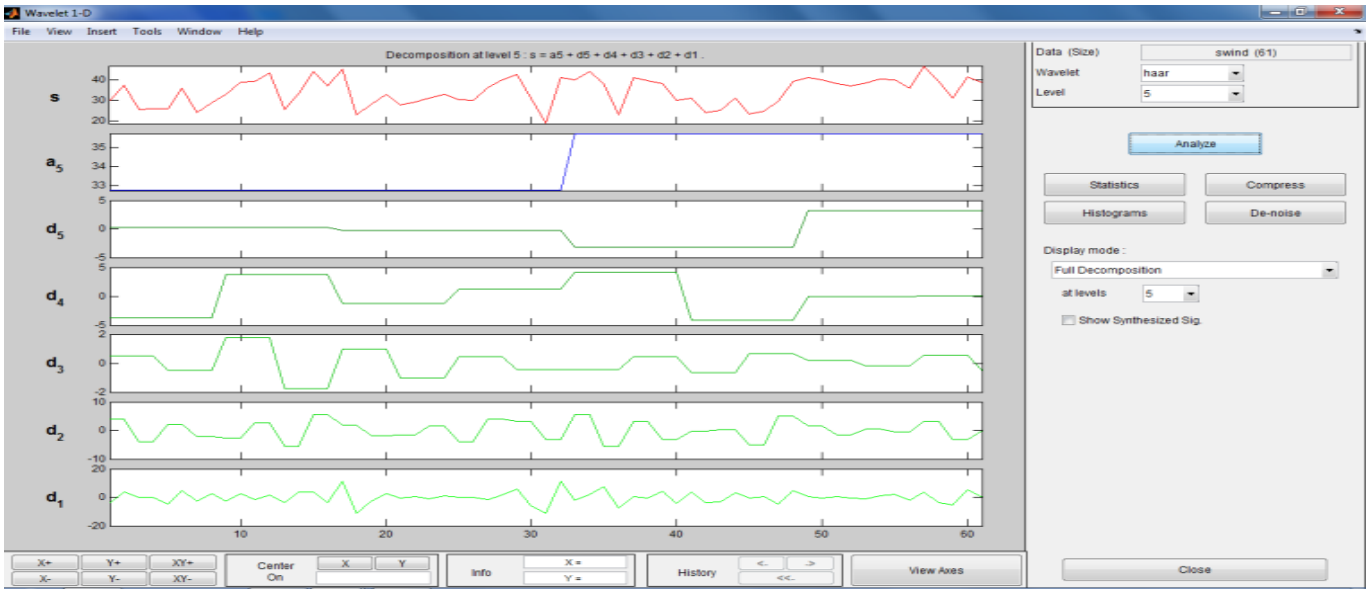

**Fig 4a:** Screen print of HAAR decomposition of Summer and Winter (right) wind speed data

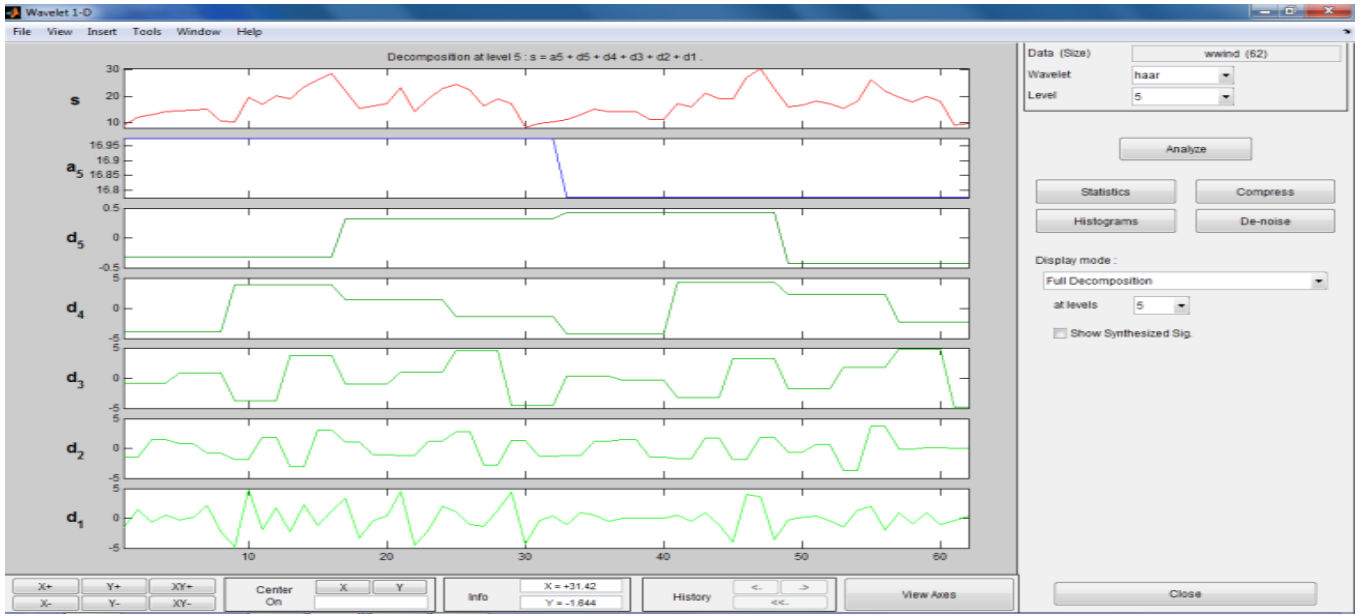

**Fig 4b:** Screen print of HAAR decomposition of Winter wind speed data

From above figure, it is clear that after 5$^{th}$ level decomposition the nature of summer and winter wind speed data clearly explains the seasonal weather difference and proves greater correlation of wind speed with weather. Therefore by HAAR transform, the summer wind data is clearly distinguished from winter wind data.

Next Morlet Continuous Wavelet Transform (CWT) is applied to get the periodicity in wind speed by taking both summer and winter wind speed data into account separately.

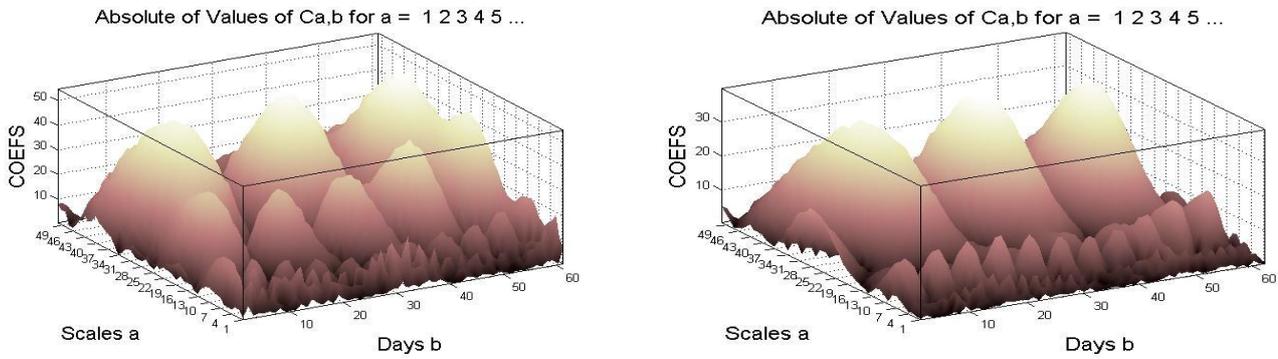

**Fig 5:** 3Dplot of CWT coefficients of summer (left) and winter (right) data

The CWT coefficient power distribution along different scales is shown by the 3D plots. In summer data periodic peaks are more than in case of winter wind data at higher scale region. At low scale periodic peaks are very low for both of the summer and winter wind data case.

The Scalograms are plotted to get better idea on scales corresponding to maximum energy or in other terms Scalograms represent percent of energy for each CWT coefficients. This is the 2D plot of the above Fig-5.

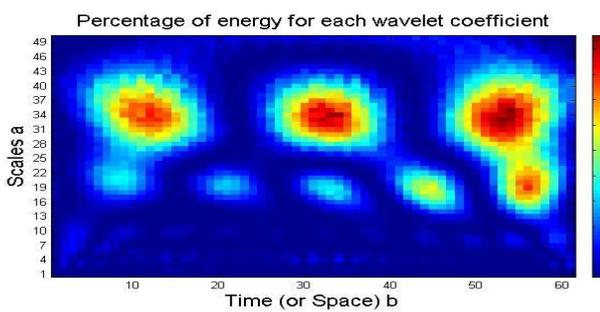 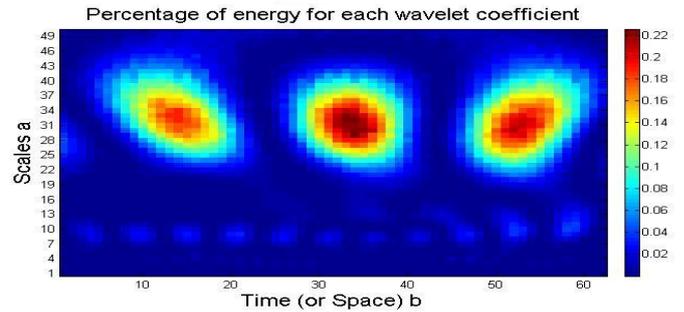

**Fig 6(a).** Scalogram of summer wind speed data          **Fig 6(b).** Scalogram of winter wind speed data

The red colour region in both the figures shows the area where maximum energy is concentrated and can be understood from the scale shown in both of the above figures. From these scalogram plots, scales 4, 9 and 32 are observed for winter data whereas scales as 4,9,19 and 34 for summer data.

The cross examination is done for the two CWT decomposed time series to understand the localized similarity in time and scale axis. In the time-frequency plane this two time series showed common power and consistent phase behavior indicating a relationship between them. To understand the similarity and difference between summer and winter wind speed where Complex Gaussian wavelet ('cgau2') is used for the clarification of indicated the phase and modulus coherence between these two.

The wavelet spectrum, defined for each signal, is characterized by the modulus and the phase of the CWT obtained using the complex-valued wavelet. The magnitude of the wavelet cross spectrum can be interpreted as the absolute value of the local covariance between the two time series in the time-scale plane.

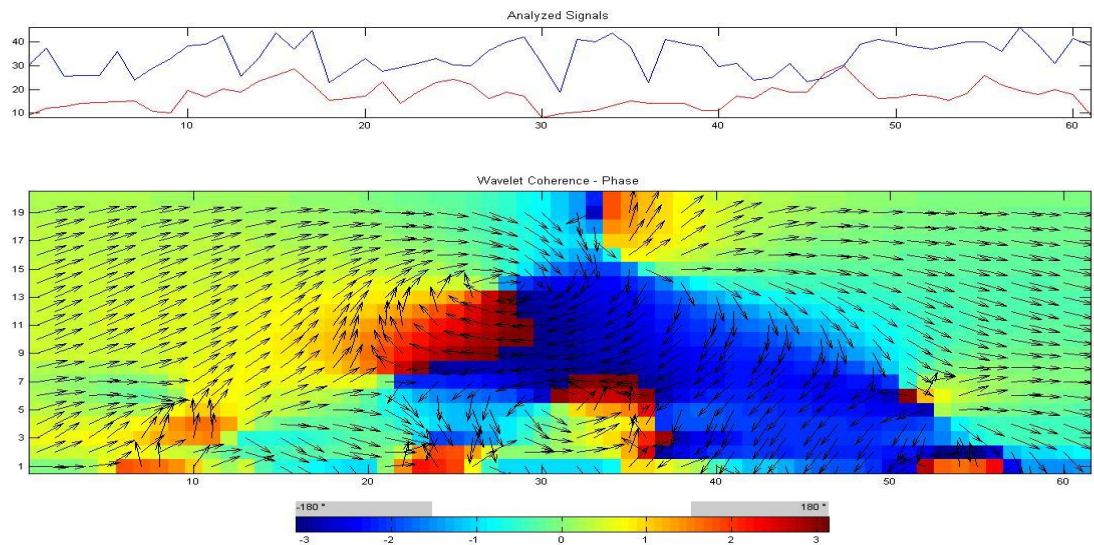

**Fig7:** Phase coherence of summer and winter time series

The arrows in the figure represent the relative phase between the two signals as a function of scale and position. The relative phase information is obtained from the smoothed estimate of the wavelet cross spectrum. The plot of the relative phases is superimposed on the wavelet coherence. The relative phase information produces a local measure of the delay between the two time series. Also the Cone of Influence (COI) can be clearly seen in the above figure. The cone consists of blue and red region showing complete out of phase relation between the two time series as indicated by the scale below the figure. This out of phase relation is understandable as the two time series are in opposite weather condition, i.e., one in summer and other in winter. So again the dependency of wind speed on season can be explained by this.

**Conclusion**

Here the comparison is made between the summer and winter wind speed to study their properties through various signal processing techniques. Using Fourier analysis the comparison is done for both the time series taking Fourier coefficients into account. Then the HAAR wavelet decomposition is used for clear distinction between these two seasonal data. Morlet wavelet is used to study the continuous nature of wind speed and by this the periodicity of wind speed is also found. At last Complex Gaussian wavelet is used to study time-frequency localization between summer and winter wind speed data and also to find out phase-modulus coherence between them. The seasonal behaviour of wind speed data of summer and winter is completely out of phase. Authors hope that the current study of the dynamics of summer and winter wind data will help the researchers move this field forward.

**References**
1. P.Bhattacharya, S.Mukhopadhyay, " Weibull Distribution for Estimating the Parameters and Application of Hilbert Transform in case of a Low Wind Speed at Kolaghat", *The International Journal of Multiphysics*, Volume-5, No-3,2011.
2. S. Mukhopadhyay,P.K.Panigrahi, A Mitra, P Bhattacharya, M Sarkar, P Das, "Optimized DHT-Neural Model as a replacement of ARMA-Neural Model for the Wind Power Forecasting purpose", *IEEE ICE International Conference Proceeding*, India, Page(s):415 – 419, 2013.
3. S.Mukhopadhyay, P.Bhattacharya, R.Bhattacharjee, P.K.Bose, " Discrete Hilbert Transform as Minimum Phase Type Filter for the Forecasting and the characterization of Wind Speed", *CODIS International IEEE Conference Proceeding*, Kolkata, India, 2012.
4. P.Bhattacharya, S.Mukhopadhyay, B.B. Ghosh, P.K.Bose, "Optimized Use of Solar Tracking System and Wind Energy", *Elsevier Procedia technology*, Volume-4, 2012, Page no- 834-839.
5. Panigrahi, P.K., Maniraman, P., Lakshmi, P.A., Yadav, R.R.: Correlations and periodicities in Himalayan tree ring widths and temperature anomalies through wavelets, http://arxiv.org/abs/nlin/0604002v1 (2006)
6. S.Mukhopadhyay, S.Mandal, P.K.Panigrahi, A.Mitra, "Heated wind particle's behavioral study by the continuous wavelet transform as well as the fractal analysis", *Computer Science & Information Technology*, pp:169-174, 2013.
7. C.P. Liu, G.S.Miller, "Wavelet Transform and Ocean current data analysis", *Journal Of Atmospheric & Ocean Technology*, Volume-3, pp:1090-1099,1996.
8. X. H. He et al.: Wavelet-Based Non-stationary Wind Speed Model in Dongting Lake Cable-Stayed Bridge, http://www.scirp.org/journal/eng (2010)
9. A.H. Siddiqi, S.Khan, S. Rehman, "Wind Speed simulation using Wavelets", *American Journal Of Applied Sciences* 2 (2), pp: 557-564, 2005.
10. S.G. Mallat, *A Wavelet Tour of Signal Processing*. 2nd ed.; Academic Press: Orlando, FL, USA, 1998.
11. S.Mukhopadhyay, P.K.Panigrahi, "Wind Speed Data Analysis for Various Seasons during a Decade by Wavelet and S transform", *International Journal in Foundations of Computer Science & Technology (IJFCST)*, Vol. 3, No.4, July 2013.
12. I. Daubechies, *Ten Lectures on Wavelets*. Philadelphia, PA: Society for Industrial and Applied Mathematics, vol. 64, 1992, CBMS-NSF Regional Conference Series in Applied Mathematics.
13. Proakis, *Digital Signal Processing*. Pearson Education India, 2007, ISBN 8131710009, 9788131710005.

**Author biographies**

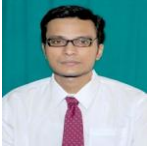
Sabyasachi Mukhopadhyay is currently pursuing MS by Research in Physical Sciences department of IISER Kolkata. Till now he has 26 numbers of International Journals, International/ National Conference Proceedings with winning the best research paper award once. His areas of research interests are Biomedical Imaging, Renewable Energy and Graph Theory. He is also the author of the author of 3books and there is an Indian patent filed in his name.

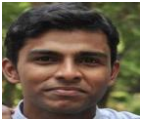
Mr. Debadatta Dash recently completed his B.Tech in Electronics & Electrical Engg. from Veer Surendra Sai University of Technology Burla, Odisha. His area of research interest is in digital signal and image processing.

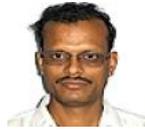
Currently Dr.Asish Mitra is the Associate Prof. in the Department of Physics of College of Engineering and Management, Kolaghat(Govt.Aided), West Bengal, India. He completed his PhD from Jadavpur University, Kolkata. He also had Post Doctorial research work from the same university. He is Reviewer of the International Journal: Heat and Mass Transfer, Publisher: Springer Berlin/Heidelberg and d WSEAS. He has several publications in reputed Journals and Conference Proceedings. His area of research interest is Computational Fluid Dynamics.

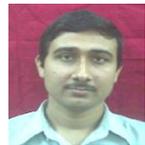
Dr. Paritosh Bhattacharya is presently working as Associate Professor in Mathematics and Head of the Department of Computer Science and Engineering at National Institute of Technology, Agartala, Tripura, India. His Twelve years of experience in higher education include his positions as a faculty member, reviewer, academic counselor, academic guide and member of academic societies. He obtained his Ph.D. (Engg) and M.Tech from Jadavpur University, West Bengal, India. He has published more than 52 papers in national and international journals and conferences. His present area of research includes Mathematical Modeling, Neural Network, Biomathematics and Computational Fluid Dynamics.